\documentclass[twocolumn,10pt]{article}

\usepackage{amsmath}
\usepackage{listings} 
\usepackage{mathrsfs} 
\usepackage{caption} 
\usepackage{balance}  
\usepackage{hyperref} 

\hypersetup
{
  colorlinks,
  linkcolor=blue,
  filecolor=blue,
  urlcolor=blue,
  citecolor=blue
}

\DeclareCaptionFormat{listingformat}{#1#2#3\hrulefill}
\newcommand{\frule}{(\mbox{$f \leftrightarrow \neg f$)}}
\newcommand{\trule}{(\mbox{$t \leftrightarrow t$)}}

\begin{document}

\title{Avoiding Contradictions in the Paradoxes, the Halting Problem, and Diagonalization}

\author{Timothy J. Armstrong\\t.armstrong888@gmail.com}

\date{}

\maketitle

\begin{abstract}

The fundamental proposal in this article is that logical formulas of the form \frule{} are not contradictions, and that formulas of the form \trule{} are not tautologies.  Such formulas, wherever they appear in mathematics, are instead reason to conclude that $f$ and $t$ have a third truth value, different from true and false.  These formulas are circular definitions of $f$ and $t$.  We can interpret the implication formula \frule{} as a rule, a procedure, to find the truth value of $f$ on the left side: we just need to find the truth value of $f$ on the right side.  When we use the rules to ask if $f$ and $t$ are true or false, we need to keep asking if they are true or false over and over, forever.

Russell's paradox and the liar paradox have the form \frule{}.  The truth value provides a straightforward means of avoiding contradictions in these problems.  One broad consequence is that the technique of proof by contradiction involving formulas of the form \frule{} becomes invalid.  One such proof by contradiction is one form of proof that the halting problem is uncomputable.  The truth value also appears in Cantor's diagonal argument, Berry's paradox, and the Grelling-Nelson paradox.

\end{abstract}

\section{Introduction}
\label{sec:intro}

Consider these Prolog rules\footnote{For an introduction to Prolog, a logic programming language, see Clocksin and Mellish \cite{clocksin-prolog-2003}.}:\\

\begin{lstlisting}
  t :- t.
  f :- \+ f.
  a(X) :- a(X).
  b(X) :- \+ b(X).
  elementOf(X, c) :- elementOf(X, X).
  elementOf(X, r) :- \+ elementOf(X, X).
\end{lstlisting}

\noindent The roughly corresponding logical formulas are:

\begin{gather}
t \leftrightarrow t\\
f \leftrightarrow \neg f\\
\forall x (A(x) \leftrightarrow A(x))\\
\forall x (B(x) \leftrightarrow \neg B(x))\\
\forall x (x \in C \leftrightarrow x \in x)\\
\forall x (x \in R \leftrightarrow x \not\in x)
\end{gather}

\noindent The Prolog predicate ``\verb|elementOf|'' is meant to be the standard ``$\in$'' symbol in set theory, and the rules involving ``\verb|elementOf|'' are meant to represent Russell's paradox. 

People familiar with Prolog should recognize that the program enters infinite recursion when we run these queries on the command line, with any constant ``\verb|z|'':\\

\begin{lstlisting}
  ?- t.
  ?- f.
  ?- a(z).
  ?- b(z).
  ?- elementOf(c, c).
  ?- elementOf(r, r).
\end{lstlisting}

\begin{sloppypar}
When we use Prolog to ask if the statements \verb|t|, \verb|f|, \verb|a(z)|, \verb|b(z)|, \verb|elementOf(c, c)|, and \verb|elementOf(r, r)| are true or false, we keep asking if they are true or false over and over, infinitely, and never arrive at an answer of true or false.  As part of the procedure to find the truth value of each statement, we need to find the truth value of the same statement.  

We should call this behavior a ``truth value''.  Some statements are true, other statements are false, and still other statements have the behavior that when we ask if they are true or false, we keep asking forever.  This sort of infinite recursion is familiar in Prolog, but we need to account for it in all forms of logic.

We would ideally like Prolog to return an answer of ``recursive'' instead of ``true'' or ``false''.  Detecting infinite recursion in general is the halting problem, but people have successfully developed algorithms to detect infinite recursion in special cases, as in the field of termination analysis \cite{termination-workshop-2014}.  There could be an option for attempting to detect infinite recursion when running a Prolog program, if it would be too computationally expensive to check for infinite recursion all the time.

This truth value is important because it appears in Russell's paradox, the liar paradox, the halting problem, Cantor's diagonal argument, Berry's paradox, and the Grelling-Nelson paradox.  Many of these problems involve formulas of the form \frule{}.  People conventionally take these formulas to be contradictions.  What Prolog's particular resolution-based theorem proving algorithm says about the statements in these problems and the above Prolog statements is that they are not true, are not false, and are not both true and false at the same time; they are not contradictions.  We should treat these statements as having the recursive truth value in all forms of logic. We need to develop a three-valued logic for this truth value; Fitting \cite{fitting-logic-1985} provides some of what is needed.

The rest of this article is organized as follows.  First it is presented how Russell's paradox has the recursive truth value.  Next, it is presented generically how \trule{} and \frule{} are recursive instead of being a tautology and a contradiction.  Afterwards, Tarski's Convention T is used to prove that the liar paradox is recursive.  The truth value has consequences for the technique of proof by contradiction, and one proof by contradiction that the halting problem is uncomputable is analyzed.  Finally, Cantor's diagonal argument is presented briefly, as well as how some numbers have this truth value in the place of some digits.  Berry's paradox and the Grelling-Nelson paradox \cite[intro]{mendelson-intro-logic-2010} are left for presentation elsewhere.
\end{sloppypar}

\section{Russell's Paradox}
\label{sec:russell}

Russell's paradox \cite{russell-types-1908} involves the set:

\begin{equation}
R = \{x \: | \: x \not\in x\}
\end{equation}

\noindent $R$ is the set of everything that is not a member of itself.  $C$ is the set of everything that \textit{is} a member of itself:

\begin{equation}
C = \{x \: | \: x \in x\}
\end{equation}

\noindent The above rules for $R$ and $C$ are repeated here:

\vspace{5px}
\begin{lstlisting}
   elementOf(X, c) :- elementOf(X, X).
   elementOf(X, r) :- \+ elementOf(X, X).
\end{lstlisting}

\vspace{0.7px}
$\forall x (x \in C \leftrightarrow x \in x)$

$\forall x (x \in R \leftrightarrow x \not\in x)$\\

For any $x$, $x$ is an element of $R$ if and only if $x$ is not an element of $x$.  We ask if $R \in R$ is true or false:

\begin{lstlisting}
   ?- elementOf(r, r).
\end{lstlisting}

\noindent Prolog enters into infinite recursion.   In the Prolog program, we provide a \textit{procedure} for determining if an arbitrary entity $x$ is an element of $R$.  In order to find out if $x$ is an element of $R$, we need to find out if $x$ is an element of itself.  In order to find out if $R$ is an element of $R$, we need to find out if $R$ is an element of itself.  

\begin{sloppypar}
It is in general desirable for sets to have a decidable procedure to determine if any entity is an element of the set.  For some sets, we can write computer programs to decide membership, as in logic programming languages like Prolog or in imperative programming languages.  We can provide logical rules so that we can use theorem proving techniques to decide membership.  It happens that in Prolog's particular theorem proving algorithm, the equivalent of \mbox{$\forall x (x \in R \leftrightarrow x \not\in x)$} becomes infinitely recursive when we use the rule to ask if $R \in R$.  Prolog interprets the implication formula as a \textit{procedure} to determine if $R \in R$:
\end{sloppypar}

\begin{equation}
R \in R \leftrightarrow R \not\in R
\end{equation}

\noindent If we can determine that $R \not\in R$, we can conclude that $R \in R$.

Russell's paradox is strange when we describe it informally: We ask if $R$ is an element of $R$.  $R$ is an element of $R$ if and only if $R$ is not an element of $R$.  In other words, $R$ is an element of $R$ if it holds that $R$ is not an element of $R$, $R$ is an element of $R$ if it is the case that $R$ is not an element of $R$, and $R$ is an element of $R$ on the condition that $R$ is not an element of $R$.  That means we have to ask: is $R$ an element of $R$?  Asking if $R$ is an element of $R$ is what we were doing at the beginning, so we ask again.  We repeat the process of asking if $R$ is an element of $R$.  When we ask if $R \in R$ is true or false, we keep asking if $R \in R$ is true or false over and over, forever.  

Is $R \in R$ \textit{actually} true or false, just we do not know?  Prolog's theorem proving algorithm leads us to conclude that $R \in R$ is neither true nor false.  Instead, it has a different truth value than true or false, the recursive truth value.  Is $R$ \textit{actually} either in $R$ or not in $R$, just we do not know?  We can never say that $R$ is either in the set or not in the set.  Instead, we keep asking forever when we ask if it is in the set.  An entity may be related to a set in a manner other than being an element of it or not an element of it.

\section{Tautologies and Contradictions}
\label{sec:tautologies}

Russell's paradox has the form of the propositional logic formula \frule{}:

\begin{gather}
(R \in R) \leftrightarrow \neg (R \in R)\\
\texttt{elementOf(r, r)} \leftrightarrow \neg \texttt{elementOf(r, r)}
\end{gather}

\begin{sloppypar}
\noindent The liar paradox, presented in section \ref{sec:liar}, also has the form \frule{}: (\mbox{$\texttt{True}(s) \leftrightarrow \neg \texttt{True}(s)$}).  The corresponding Prolog rule is:
\end{sloppypar}

\begin{lstlisting}
   f :- \+ f.
\end{lstlisting}

$f$ is true if and only if $f$ is false.  $f$ is true if it holds that $f$ is false; $f$ is true on the condition that $f$ is false.  The Prolog rule provides a means, a procedure, for finding the truth value of $f$.  In general, if we want to find the truth value of $f$, we need to search the Prolog database to find if there is a fact asserting $f$, or if there is a rule with $f$ as its head.  We find the rule ``\verb|f :- \+ f.|'', and we attempt to satisfy the body of the rule.  As part of the procedure to find the truth value of $f$, we need to find the truth value of $f$.  When we ask if $f$ is true or false, we keep asking repeatedly forever.

In classical two-valued logic, we often interpret an implication statement \mbox{($p \leftrightarrow q$)} as providing a procedure, a rule, for finding the truth value of $p$: we just need to find the truth value of $q$.  This interpretation of implication is explicit in Prolog.  We should interpret \frule{} as providing a rule for finding the truth value of $f$ on the left side: we just need to find the truth value of $f$ on the right side.

If we somehow know that $f$ is either true or false, the formula \frule{} would force us to conclude that $f$ has the opposite truth value, which would be a contradiction.  However, if all we have is the formula \frule{}, it is just a rule for finding the truth value of $f$: an infinitely recursive rule.  $f$ is true if and only if $f$ is false.  So, using this rule, in order to find out if $f$ is true, we need to find out if $f$ is false.

Also consider the formula \trule{} and the Prolog rule ``\verb|t :- t.|''.  As part of the procedure to find the truth value of $t$, we need to find the truth value of $t$.  The set $C$ above has the form \trule{}:

\begin{gather}
(C \in C) \leftrightarrow (C \in C)\\
\texttt{elementOf(c, c)} \leftrightarrow \texttt{elementOf(c, c)}
\end{gather}

\noindent As part of the procedure to find out if $C$ is an element of $C$, we need to find out if $C$ is an element of $C$.

In classical two-valued logic, \frule{} is a contradiction, and \trule{} is a tautology.  The proposal in this article is that we should instead treat these formulas as reason to conclude that $f$ and $t$ have the recursive truth value.  $f$ and $t$ have just a \textit{single} truth value; they are not true, are not false, and are not both true and false at the same time.  We should treat all formulas with the \textit{form} \frule{} or \trule{} as being infinitely recursive, such as Russell's paradox and the liar paradox.

It makes sense to say that \trule{} is a tautology in that, if we know that $t$ is true, we can conclude that $t$ is true.  On the other hand, if we intend \trule{}  to be a \textit{rule} for finding the truth value of $t$, as in Prolog, then we would say that \trule{} is not a tautology but instead says something special about the truth value of $t$, that $t$ has the recursive truth value.

Saying that \frule{} is not a contradiction seems like a bold claim.  For one matter, it would invalidate one form of the technique of proof by contradiction.  We should say that it is still a contradiction if a statement has more than one truth value at the same time, such as ($p \wedge \neg p$); proofs by contradiction of that sort would still be valid.  However, proofs by contradiction that depend on a formula of the form \frule{} being a contradiction would be invalid.  We would need to sort through all of mathematics to find all the proofs by contradiction that have this form, and figure out how to correct the proofs and all the theory built on top of those proofs.  It would be a very large task.

Formulas of the form \frule{} would be legitimate to have as axioms or theorems in a formal theory, or as data in a knowledge base, and would not make the theory or knowledge base inconsistent.  That observation is consequential for the paradoxes.  Formulas of the form \trule{} would not be \textit{harmless} to have in a theory or knowledge base.

\section{The Liar Paradox}
\label{sec:liar}

It may be evident from what was presented above that the liar paradox has the recursive truth value.  There is much more that needs to be said about the liar paradox, which is not included in this article for space considerations.  Let us briefly consider, though, Tarski's well-known ``Convention T'' \cite{tarski-semantic-1944}.  He writes:

\begin{quote}
Let us consider an arbitrary sentence; we shall replace it by the letter `$p$.'  We form the name of this sentence and we replace it by another letter, say `$X$.'  We ask now what is the logical relation between the two sentences ``$X$ \textit{is true}'' and `$p$.'  It is clear that from the point of view of our basic conception of truth these sentences are equivalent. In other words, the following equivalence holds:

\vspace{5px}
(T) \textit{X is true if, and only if, p.}
\end{quote}

\noindent Tarski provides the example of the sentence ``snow is white'':

\begin{quote}
The sentence ``snow is white'' is true if, and only if, snow is white.
\end{quote}

For the liar paradox, let us represent with the letter `$s$' the sentence ``This sentence is not true'', or equivalently, ``Sentence `$s$' is not true''.  Then, by Convention T: ``Sentence `$s$' is true if, and only if, sentence `$s$' is not true.''  We can formalize the formula roughly as:

\begin{equation}
\verb|True|(s) \leftrightarrow \neg \verb|True|(s)
\end{equation}

\noindent The formula has the form \frule{}.  ``\verb|True|'' is the truth predicate that asserts that the argument is a true sentence.  What we want to say in connection to the recursive truth value is that Convention T provides a bidirectional rule: if we know that $p$, we can conclude that $X$ is true; if we know that $X$ is true, we can conclude that $p$.  The most \textit{direct} way to find out if $X$ is true is to find out if $p$.  The most direct way to find out if the sentence ``snow is white'' is true is to find out if snow is white.  The most direct way to find out if sentence `$s$' is true is to find out if `$s$' is not true.  

Convention T provides a \textit{procedure} for finding out if $X$ is true: we need to find out if $p$.  Using this rule, in order to find out if `$s$' is true, we need to find out if `$s$' is true.  In order to find the truth value of `$s$', we need to find out if its claim about reality is correct.

`$s$' thus has the recursive truth value.  `$s$' is true if it is not true and is not true if it is true; so, when we ask if `$s$' is true, we need to ask again repeatedly forever.   We initially do not know the truth value of `$s$'.  We need some means, some procedure, for finding its truth value.  Convention T provides such a procedure.

It is similar for ``This sentence is true'', which is true if and only if it is true.  If we call that sentence `$u$', we can write:

\begin{equation}
\verb|True|(u) \leftrightarrow \verb|True|(u)
\end{equation}

\noindent which has the form \trule{}.  As part of the process to find out if `$u$' is true, we need to find out if `$u$' is true.  (We would need to consider in more detail elsewhere the difference between `$s$' and ``This sentence is false'', but in any case both are infinitely recursive.)

\section{The Halting Problem as a Proof by Contradiction}
\label{sec:halting}

One notable proof by contradiction with the form \frule{} is the proof by Davis et al. that the halting problem is uncomputable \cite[ch 4]{davis-computability-1994}.  Turing's proof is a bit different, but Davis et al.'s proof is explicitly in this form.  The authors assume that the halting problem is computable and then arrive at this formula that they claim is a contradiction:

\begin{equation}
\verb|HALT|(y_0, y_0) \leftrightarrow \neg \, \verb|HALT|(y_0, y_0)
\label{eqn:halt-contradiction}
\end{equation}

\noindent They conclude, by a proof by contradiction, that their assumption that halting problem is computable must be false.

Equation \ref{eqn:halt-contradiction} has the form \frule{}.  It makes sense, given how the authors present the halting problem, to say that \verb|HALT|($y_0, y_0$) has the recursive truth value: to say that, in order to find the truth value of \verb|HALT|($y_0, y_0$) on the left side, we need to find the truth value of \verb|HALT|($y_0, y_0$) on the right side.

It is worth examining in a bit more detail.  The authors discuss computability mainly using an imperative programming language they devised, instead of using Turing machines.  \verb|HALT()| is a computer program that takes as its first argument a natural number $x$, and as its second argument a natural number $y$ representing an arbitrary computer program, and is supposed to decide if the program $y$ running with the input of $x$ would either halt or run forever (on an idealized computer).  \verb|HALT()| provides a return value of ``\verb|true|'' (or ``1'') if program $y$ would halt and ``\verb|false|'' (or ``0'') if it would not halt.  \verb|HALT()| is supposed to compute the function:

\begin{equation}
\texttt{HALT}(x, y) = \left\{ 
 \begin{array}{l l}
 	1 & \quad \text{if program $y$ running}\\
 	  & \:\:\:\:\:\:\:\:\:\text{with input $x$ halts}\\
 	0 & \quad \text{otherwise}
 \end{array} \right.
 \end{equation}

The authors construct a certain program $\mathscr{P}$ that is problematic:\\

\verb|[A] IF HALT(X, X) GOTO A|\\

\noindent $\mathscr{P}$ translated into C/C++ syntax\footnote{For an introduction to C, an imperative programming language, see Kernighan and Ritchie \cite{kernighan-c-1988}.} (with which the reader may be more familiar) is in listing \ref{lst:halting-program}.  $y_0$ is the natural number that represents $\mathscr{P}$.

\begin{lstlisting}[float=t,frame=tb,language=C++,label=lst:halting-program,caption={The program $\mathscr{P}$ that is problematic for the halting problem, given in C/C++ syntax.  To explain the code for readers unfamilar with C/C++: The first \mbox{``\texttt{unsigned int}''} means that the return value of the program is a natural number.  The second \mbox{``\texttt{unsigned int}''} means that the program takes a single parameter ``\texttt{x}'' that is a natural number.  ``\texttt{A:}'' is a label for the given line.  If the \texttt{HALT(x, x)} procedure call evaluates to ``\texttt{true}'', the ``\texttt{goto A}'' command causes the program to enter into an infinite loop, repeatedly executing line \texttt{A}.  If the \texttt{HALT(x, x)} procedure call evaluates to ``\texttt{false}'', the program reaches the ``\texttt{return 0;}'' command, which halts the program and returns the answer of 0.}]
unsigned int P(unsigned int x)
{
A:  if (HALT(x, x)) goto A;

    return 0;
}
\end{lstlisting}

Consider what happens when we run $\mathscr{P}$ with the input of $y_0$, that is when we run $\mathscr{P}(y_0)$.  If, inside $\mathscr{P}$, the \verb|HALT|($y_0, y_0$) procedure call returns ``\verb|true|'' (saying $\mathscr{P}(y_0)$ halts), then $\mathscr{P}(y_0)$ does not halt.    If the \verb|HALT|($y_0, y_0$) procedure call returns ``\verb|false|'' (saying $\mathscr{P}(y_0)$ does not halt), then $\mathscr{P}(y_0)$ halts.  Thus, as in equation \ref{eqn:halt-contradiction}:\\

\verb|HALT|($y_0, y_0$) $\leftrightarrow \neg $\verb|HALT|($y_0, y_0$)\\

What we want to say here about the recursive truth value is as follows.  In order to find out if $\mathscr{P}(y_0)$ halts (that is, in order to find the truth value of \verb|HALT|($y_0, y_0$)), we need to find the return value of the \verb|HALT|($y_0, y_0$) procedure call inside $\mathscr{P}$.  That is, in order for us to find the truth value of \verb|HALT|($y_0, y_0$), we need the \verb|HALT()| program to tell us the truth value of \verb|HALT|($y_0, y_0$).

In this interpretation, \verb|HALT|($y_0, y_0$) has the recursive truth value.  When we ask if \verb|HALT|($y_0, y_0$) is true or false, we -- or the \verb|HALT()| program -- need to keep asking if \verb|HALT|($y_0, y_0$) is true or false over and over, forever.

It is simplest to say that Davis et al.'s proof by contradiction, asserting that the halting problem is uncomputable, is invalid because equation \ref{eqn:halt-contradiction}, having the form \frule{}, is not actually a contradiction.

\section{Diagonalization and the Halting Problem}
\label{sec:diagonalization}

The reader may be able to imagine how the recursive truth value relates to Cantor's diagonal argument \cite{cantor-diagonalization-1892} and to Turing's original article on his version of the halting problem \cite{turing-computable-1937}.  These topics require more extended presentation, but we should say a few words briefly here.

Some numbers have the recursive truth value in the place of some digits.  Say we have written a computer program to perform the computation of finding the digits of a real number.  For some numbers and for some digits, when we attempt to find the value of the digit, we need to attempt again to find the value of the same digit.  The program enters infinite recursion.  When we ask what the value of the digit is, we keep asking what the value is over and over, forever.  For one example, if we interpret the set in Russell's paradox as a real number, it has an infinitely recursive digit.

If it is possible to detect infinite recursion, though, the program can just mark the given digit as having the recursive truth value, such as with an ``r'', and move on to computing the next digit.  For example, we could write: ``0.10r0110...''

There is a number with an infinitely recursive digit in Turing's article in section 8, ``Application of the diagonal process'', which is the key section for the halting problem and the recursive truth value.  In Turing's article, in order to find the $R(K)$-th digit of $\beta^\prime$, the machine needs to find the $R(K)$-th digit of $\beta^\prime$.  However, the recursive truth value allows us to handle such a number.  What the machine can do is simply mark the $R(K)$-th digit as having the recursive truth value, such as by printing an ``r'' on the tape, and move on to computing the next digit in $\beta^\prime$.  In this way, the machine running with its own program number as input becomes less problematic.

For Cantor's diagonal argument, it happens that, when we attempt to include the diagonal and anti-diagonal real numbers as rows in the matrix, the numbers acquire an infinitely recursive digit on the diagonal.  A proof requires more extended presentation, but let us just comment on a mathematical equation that Boolos et al. use to explain diagonalization \cite[ch 2]{boolos-computability-2007}.  They assert that this equation is a contradiction:

\begin{equation}
s_m(m) = 1 - s_m(m)
\end{equation}

\noindent $s_m(m)$ is supposed to take the value of either 0 or 1.  We should treat this equation as being infinitely recursive instead of as a contradiction: in order to find the value of $s_m(m)$ on the left, we need to find the value of $s_m(m)$ on the right.  We define the value of $s_m(m)$ to be 1 if the value of $s_m(m)$ is 0, and to be 0 if the value of $s_m(m)$ is 1.  

There is a similar equation in Turing's article in section 8, which we can rearrange to:

\begin{equation}
\phi_K(K) = 1 - \phi_K(K)
\end{equation}

\noindent Turing asserts that this equation is a contradiction, but we should instead treat it as infinitely recursive: in order to find the value of $\phi_K(K)$ on the left, we need to find the value of $\phi_K(K)$ on the right.

\section{Conclusion}
\label{sec:conclusion}

The reader should hopefully find it plausible, and perhaps convincing, that it is best to treat \frule{} as infinitely recursive instead of as a contradiction.  It seems very clear in Prolog that \trule{} and \frule{} lead us to conclude that $t$ and $f$ have the recursive truth value.  It provides a convenient means of avoiding contradictions in the paradoxes, the halting problem, and diagonalization.  This approach to handling the paradoxes provides an alternative to Zermelo-Fraenkel set theory, type theory, and Tarski's hierarchy of languages.  We would need to re-work the foundations of logic and mathematics to include this truth value.  

We can consider how the truth value works in propositional logic with \trule{} and \frule{}, before considering how it works in first-order logic and other forms of logic.  It would be necessary to figure out for all the proof systems (truth tables, resolution, tableaux, axiomatic systems, etc.) how to prevent them from proving that \trule{} is a tautology, and how to prevent them from proving that \frule{} is a contradiction.  It would be necessary to figure out how to adapt the proof systems so that, when given \trule{} and \frule{} as premises, they prove that $t$ and $f$ have the recursive truth value.

\balance 


\begin{thebibliography}{10}

\bibitem{termination-workshop-2014}
{\em 14th International Workshop on Termination}, 2014.

\bibitem{boolos-computability-2007}
George~S Boolos, John~P Burgess, and Richard~C Jeffrey.
\newblock {\em Computability and Logic}.
\newblock Cambridge University Press, fifth edition, 2007.

\bibitem{cantor-diagonalization-1892}
Georg Cantor.
\newblock Ueber eine elementare {F}rage der {M}annigfaltigkeitslehre.
\newblock {\em Jahresbericht der Deutschen Mathematiker-Vereinigung}, 1:75--78,
  1892.

\bibitem{clocksin-prolog-2003}
William~F Clocksin and Christopher~S Mellish.
\newblock {\em Programming in Prolog: Using the ISO Standard}.
\newblock Springer, fifth edition, 2003.

\bibitem{davis-computability-1994}
Martin~D Davis, Ron Sigal, and Elaine~J Weyuker.
\newblock {\em Computability, Complexity, and Languages: Fundamentals of
  Theoretical Computer Science}.
\newblock Academic Press, second edition, 1994.

\bibitem{fitting-logic-1985}
Melvin Fitting.
\newblock A {K}ripke-{K}leene semantics for logic programs.
\newblock {\em The Journal of Logic Programming}, 2(4):295--312, 1985.

\bibitem{kernighan-c-1988}
Brian~W Kernighan and Dennis~M Ritchie.
\newblock {\em The {C} Programming Language}.
\newblock Prentice-Hall, second edition, 1988.

\bibitem{mendelson-intro-logic-2010}
Elliott Mendelson.
\newblock {\em Introduction to Mathematical Logic}.
\newblock CRC Press, fifth edition, 2010.

\bibitem{russell-types-1908}
Bertrand Russell.
\newblock Mathematical logic as based on the theory of types.
\newblock {\em American Journal of Mathematics}, 30(3):222--262, 1908.

\bibitem{tarski-semantic-1944}
Alfred Tarski.
\newblock The semantic conception of truth: {A}nd the foundations of semantics.
\newblock {\em Philosophy and Phenomenological Research}, 4(3):341--376, 1944.

\bibitem{turing-computable-1937}
Alan~M Turing.
\newblock On computable numbers, with an application to the
  {E}ntscheidungsproblem.
\newblock {\em Proceedings of the London Mathematical Society}, 42:230--265,
  1937.

\end{thebibliography}
\end{document}